# NORMALIZATION OF REGRESSOR EXCITATION AS A PART OF DYNAMIC REGRESSOR EXTENSION AND MIXING PROCEDURE[1]

**A. Glushchenko, *Member*, *IEEE*, V. Petrov, K. Lastochkin**
Stary Oskol technological institute n.a. A.A. Ugarov (branch) NUST "MISIS"

a.glushchenko@sf-misis.ru

The method of excitation normalization of the regressor, which is used in the estimation loop to solve the plant identification problem, is proposed. It is based on the dynamic regressor extension and mixing procedure. Its application allows to obtain the same upper bound of the parameter identification error for the scalar regressors with different excitation level, using a constant value of the adaptation rate for all of them. This fact is a significant advantage from the practical point of view. Comparison of the developed method with the known one of the regressor amplitude normalization is conducted. It is shown that the classical approach does not have the above-stated property. To validate the theoretical conclusions made, the results of the comparative mathematical modeling of three loops are presented: 1) the classical gradient one, 2) the one with the normalization of the regressor amplitude, 3) the proposed one with the normalization of the regressor excitation.

*Keywords:* identification, gradient descent, parameter error, adaptation rate of parameter estimation loop, regressor excitation level, normalization.

## 1. Introduction

Considering various branches of industry, the methods of classical theory of the dynamical plants identification [1] are widespread in technical practice of engineers of technological processes automation. As a rule, they are applied in the course of commissioning to obtain a mathematical model of a technological unit. Such model is then usually used to calculate the parameters of the widespread PI and PID controllers. The accuracy of the model is a critical parameter because it directly determines the accuracy of the calculated controller parameters and, therefore, the economic indicators of the technological unit, which, in turn, depend on the control quality.

The classical methods of the identification theory, which, first of all, include the recursive least-squares and the gradient descent ones, provide accurate identification of the model parameters only if the condition of the regressor persistent excitation is satisfied [2]. In practice, this requires injection of a high-frequency test signal into the control input of the plant [3]. It is not always possible to meet this requirement under the real industrial production conditions. Therefore, to relax the condition of the persistent excitation, new methods [4-9] have been proposed in the literature over the last few years. They somehow improve the model parameter identification accuracy and, consequently, its adequacy. The main obtained results are the new estimation loops with the finite-time convergence [4-6] and various filtering and regressor preprocessing algorithms [7-10]. They allow one to avoid using the test signal injection procedure and identify the model parameters with high accuracy when the technological unit functions in an "almost-normal" mode. In general, with some limitations, the problem of accurate identification of the time-invariant parameters of the dynamical plants can be considered to be well-developed with fully clear further research lines and promising practical applications in industry [11, 12].

[1]Research was financially supported by Russian Foundation for Basic Research (Grant 18-47-310003-r_a).

Among the actual lines of research, the transfer of the obtained results to the regression equations with nonlinear parameterization [13] and step-like changing or nonstationary plant parameters [6, 9,14] should be noted.

Another, less studied problem of both classical and modern identification (parameter estimation) loops is the choice of the value of the adaptation rate $\gamma$ of the adaptation (identification) law [15]. This problem is even more complex for the identification loops, which provide the finite-time convergence of the model parameters estimates [4-6]. In such case the value of $\gamma$ is to be chosen taking into consideration some special conditions.

Considering the classical gradient based identification loop, as well as the gradient-type ones with the regressor preprocessing procedures [7-10], this problem can be divided into the technical and theoretical levels.

As far as the technical level is concerned, the problem is that it is impossible to solve numerically the stiff differential equation of the estimation loop, which is obtained when the high value of the adaptation rate is used, by means of the modern industrial microcontrollers [15].

As for the theoretical level, the above-mentioned problem lies in: a) the fact that the convergence rate of the model parameter estimates depends on the adaptation rate value gain $\gamma$ directly, while the accuracy of such estimates obtained in case of disturbances depends on $\gamma$ inversely, and b) different values of $\gamma$ should be used depending on the regressor excitation level value [1].

In this paper, the point (b) of the theoretical level of the $\gamma$ value choice problem is considered in more detail for the gradient-type identification laws.

The main difficulty in choice of the adaptation rate value for both the basic gradient estimation loop and its improved analogues, which are based on the regressor preprocessing [7-10], consists in the dependence between the upper bound of the parameter error and the product of the estimation loop adaptation rate $\gamma$ by the level of the regressor excitation $\alpha$. It is necessary to note here that, considering the conventional gradient estimation loop, the above-mentioned dependence exists under some assumptions [16]. At the same time, this dependence takes place without any assumptions for most modern estimation loops with regressor preprocessing.

The value of the level of the regressor excitation $\alpha$ is generally defined by the excitation interval length $T$ and the amplitude $A$ of the regressor. So, the above-mentioned dependence implies that the upper bound of the parameter error takes on different values for the regressors with different levels of excitation. In other words, when the value of $\gamma$ is constant, the parameter error can be extremely high for some regressors and, on the contrary, extremely low for others. From the point of view of the practical industrial application of the modern methods of the identification theory, it would be highly advisable to have the same upper bound of the parameter error for the regressors with different levels of excitation $\alpha$.

For this purpose, the following methods have been proposed in the literature: 1) the scaling method [17], which allows to keep approximately the same upper bound of the parameter error by changing the estimation loop adaptation rate $\gamma$ depending on the indirect estimation of the regressor excitation level $\alpha$; 2) the recursive least squares method with exponential forgetting [14, 15, 17], which provides the dependence of the upper bound of the parameter error only on the exponential forgetting factor in the limit. The scaling method requires manual selection of a number of scaling parameters, while the recursive least squares one provides the required property of proportionality of the upper bound of the parameter error and the forgetting factor only in the limit and only for the regressors of a certain kind.

The scaling and the recursive least-squares with exponential forgetting methods are based on the general idea of keeping the value of γα constant with the help of the online adjustment of the adaptation rate γ depending on the estimate of the regressor excitation level α. However, according to the viewpoint of the authors of this research, such approach, in the end, inevitably leads either to a new adaptation loop with the new parameters, which require to be chosen manually, or to application of various kinds of heuristics.

Therefore, in this paper, in order to obtain the same upper bound of the parameter error for the regressors of different excitation level, it is proposed not to adjust the adaptation rate γ, but to develop a procedure of normalization of the regressor excitation. It will allow one to obtain the dependence of the parameter error not on the value of γα, but on a new product γΔ, where Δ is a value, which is, unlike α, independent of the regressor amplitude $A$.

The proposed method will allow one to use a constant adaptation rate γ and still have the same upper bound of the parameter error for different regressors.

It is proposed to develop an estimation loop, which has the above-stated properties, by modifying one of the known regressor preprocessing methods – the dynamic regressor extension and mixing procedure [10].

## 2. Formal description and statement of the problem

The problem of the parameters identification for a class of linear plants is considered:

$$y(t) = \frac{b(p)}{a(p)} u(t), \quad (2.1)$$

where $p = d/dt$ is the differentiation operator, $y$ is an output variable, $u$ is a control action signal, $b(p) = \sum_{i=0}^{m} b_i p^i$ and $a(p) = p^n + \sum_{i=0}^{n-1} a_i p^i$ are the polynomials with the quasi-steady ($\dot{b}_i \approx 0$, $\dot{a}_i \approx 0$) unknown parameters.

Model (2.1) can be written as a linear regression using the well-known method [15]:

$$\bar{z}(t) = \theta^T \bar{\omega}(t) = y(t) + \psi^T \bar{\omega}_2(t),$$

$$\theta = [b_m, b_{m-1}, \ldots, b_0, a_{n-1}, a_{n-2}, \ldots, a_0]^T, \quad (2.2)$$

$$\bar{\omega}(t) = \left[ \frac{\alpha_m(p)}{\Psi(p)} u(t) \quad -\frac{\alpha_{n-1}(p)}{\Psi(p)} y(t) \right]^T = \left[ \bar{\omega}_1^T(t) \quad \bar{\omega}_2^T(t) \right]^T.$$

where $\bar{z} \in R$ is a measurable function, $\bar{\omega} \in R^{m+n+1}$ is a measurable regressor, θ is a vector of the unknown parameters, $\alpha_i(p) = [p^i, p^{i-1}, \ldots, 1]^T$ is the differentiation operator, $\Psi(p) = p^n + \psi^T \alpha_{n-1}(p)$ is a stable polynomial with $\psi = [\psi_{n-1}, \psi_{n-2}, \ldots, \psi_0]^T$.

Let the procedure of the dynamic regressor extension and mixing (DREM) [10] be applied to the regression (2.2). For this purpose, $m+n$ delay operators are introduced:

$$(.)_{f_i(t)} := [H_i(.)](t) = (.)(t - d_i); \; i \in \{1, 2, \cdots, m+n\}, \quad (2.3)$$

where $d_i$ is the delay value.

The extended linear regression equation is obtained as a result of application of the filter (2.3) to the function $\bar{z}$ and regressor $\bar{\omega}$:

$$\bar{z}_f(t) = \bar{\omega}_f(t) \theta$$

$$\bar{z}_f(t) = \left[ \bar{z}(t) \; \bar{z}_{f_1}(t) \; \ldots \; \bar{z}_{f_{n+m}}(t) \right]^T; \; \bar{\omega}_f(t) = \left[ \bar{\omega}(t) \; \bar{\omega}_{f_1}(t) \; \ldots \; \bar{\omega}_{f_{n+m}}(t) \right]^T. \quad (2.4)$$

Having multiplied the equation (2.4) from the left by the adjoint matrix $\mathrm{adj}\{\overline{\omega}_f(t)\}$ of the extended regressor $\overline{\omega}_f(t)$ and used the equality $\mathrm{adj}\{\overline{\omega}_f(t)\}\overline{\omega}_f(t) = \det\{\overline{\omega}_f(t)\}I$, the following equation is obtained:

$$z(t) = \mathrm{adj}\{\overline{\omega}_f(t)\}\overline{z}_f(t) = \mathrm{adj}\{\overline{\omega}_f(t)\}\overline{\omega}_f(t)\theta = \det\{\overline{\omega}_f(t)\}\theta = \omega(t)\theta. \quad (2.5)$$

where $z \in R^{m+n+1}$, $\omega \in R$.

Using the conventional gradient descent method, the estimation loop for the regression (2.5) is written as:

$$\dot{\tilde{\theta}}_i(t) = \dot{\hat{\theta}}_i(t) = -\gamma\omega\left(\hat{\theta}_i\omega - z\right) = -\gamma\omega^2\tilde{\theta}_i(t), \quad (2.6)$$

Then the differential equation (2.6) is solved:

$$\tilde{\theta}_i(t) = e^{-\gamma\int_{t_0}^{t}\omega^2(\tau)d\tau}\tilde{\theta}_i(t_0). \quad (2.7)$$

where $t_0$ is the moment of time when the output of the operator (2.3) with the maximal $d_i$ value becomes non-zero.

Let the notion of the excitation level of the scalar regressor over time range $[t_s; t_s+T]$ be introduced:

***Definition 1:*** *The regressor $\omega$ is finitely exciting ($\omega \in$ FE) over the interval $[t_s; t_s+T]$ if there exists $t_s \geq t_0$, $T > 0$ and $\alpha > 0$ such that the following inequality holds:*

$$\int_{t_s}^{t_s+T} \omega^2(\tau)d\tau \geq \alpha, \quad (2.8)$$

*where $\alpha > 0$ is the excitation level.*

Taking into consideration (2.8), the parameter error equation (2.7) is rewritten over the interval $[t_s; t_s+T]$:

$$\tilde{\theta}_i(t_s + T) = e^{-\gamma\int_{t_s}^{t_s+T}\omega^2 d\tau}\tilde{\theta}_i(t_s) \leq e^{-\gamma\alpha}\tilde{\theta}_i(t_s) < \tilde{\theta}_i(t_s). \quad (2.9)$$

As it follows from (2.9) and (2.8), the parameter error $\tilde{\theta}_i(t_s + T)$ is bounded from above by the expression, which depends on the multiplication $\gamma\alpha$. From the point of view of the practical application of the estimation loop (2.7), it is advisable to keep the same identification accuracy (i.e., the same upper bound of the parameter error (2.9)) for different regressors. However, as it follows from (2.8), the value of $\alpha$ is different for different regressors and defined by both the length $T$ of the excitation interval and the amplitude $A$ of the regressor. So, to keep the same value of $\gamma\alpha$ when the parameters $A$ and $T$ are time-varying, the adaptation rate $\gamma$ of the estimation loop (2.7) is to be adjusted. Otherwise, when $\gamma$ is constant, if $\alpha \to 0$ and $\alpha > 0$, then $\tilde{\theta}_i(t_s + T) \to \tilde{\theta}_i(t_s)$, but if $\alpha \to \infty$, then, contrary, $\tilde{\theta}_i(t_s + T) \to 0$.

The aim of this research is to solve the problem of keeping the same upper bound (2.9) of the parameter error (2.7) for a class of the regressors with various values of $A$, but which are exciting over the intervals of the same length $T$.

***Definition 2:*** *The regressors $\omega_j$ are finitely exciting over the same time interval $[t_s; t_s+T]$ if there exists $t_s \geq t_0$, $T > 0$ and $\alpha_j > 0$ such that the following inequalities hold true:*

$$\int_{t_s}^{t_s+T} \omega_j^2(\tau)d\tau \geq \alpha_j, \quad (2.10)$$

*where $\alpha_j$ is the excitation level of the $j^{th}$ regressor.*

***Remark 1.*** *Particularly, the regressors, which are obtained as a result of the application of the procedure (2.2)-(2.5) using the control action of different amplitudes (e.g. u = 1, u = 10, u = 100), belong to the class (2.10).*

On a particular numerical example, let the need to adjust the adaptation rate $\gamma$ for regressors from the class (2.10) be demonstrated.

***Example 1.*** Let $t_s = 0$ s, $T = 10$ s. Then, when $\omega = Ae^{-1t}$ in (2.9), it is obtained:

$$\tilde{\theta}_i(10) = e^{-\gamma A^2 \int_0^{10} e^{-2t} d\tau} \tilde{\theta}_i(0) \approx e^{-\gamma A^2 0.5} \tilde{\theta}_i(0), \quad (2.11)$$

Let $A$ tend to zero and $\gamma$ be constant. As a result, using (2.11), $\tilde{\theta}_i(10) \to \tilde{\theta}_i(0)$. This means that: 1) *the upper bound of the parameter error varies for different regressors, which belong to the class* (2.10); 2) *if the regressor amplitude changes its value, then the adaptation rate $\gamma$ is to be chosen again (or adjusted) to keep the same upper bound of the parameter error.*

It would be possible to avoid the necessity to chose $\gamma$ value again and again by having some normalized regressor $\varphi = \omega f(\omega)$ instead of $\omega$ for the estimation loop (2.6)-(2.9). For any $\varphi$, the following inequality must be true:

$$\int_{t_s}^{t_s+T} \varphi^2(\tau) d\tau \geq \Delta > 0. \quad (2.12)$$

Then the parameter $\gamma$ will not need to be adjusted, and the same upper bound of the parameter error will be provided for the regressors from the class (2.10).

Thus, the aim of this research is to develop a normalization function $f(\omega)$ that allows one to obtain the same upper bound of the parameter error for the regressors from class (2.10) under the condition that $\gamma$ is chosen once and then kept constant.

***Remark 2.*** T*he exponentially decaying regressor is used in Example* 1 *for illustration only. The above-demonstrated problem is typical for all regressors of different amplitudes, which are exciting over the intervals of the same length T.*

## 3. Main result

Let the regressor $\omega$ be presented in the numerical notation form:

$$\omega = \text{sgn}(\omega) 10^\eta, \quad (3.1)$$

where $\eta$ and the sign function $\text{sgn}(\omega)$ are defined as follows:

$$\text{sgn}(\omega) = \begin{cases} 1 & \text{if } \omega \geq 0 \\ -1 & \text{otherwise} \end{cases}; \quad \eta = \begin{cases} \log_{10}(|\omega|) & \text{if } |\omega| \neq 0 \approx 10^{-\infty} \\ -\infty & \text{otherwise} \end{cases}. \quad (3.2)$$

Using the definitions (3.1) и (3.2), the normalization function is introduced:

$$f(\omega) = \text{sgn}(\omega) 10^{-sat(\eta)}, \quad (3.3)$$

where $sat(\eta)$ is:

$$\text{sat}(\eta) = \begin{cases} \eta_{\min} & \text{if } \eta \leq \eta_{\min} \\ \eta & \text{otherwise} \end{cases}. \quad (3.4)$$

Let the regression (2.5) be multiplied by the normalization function (3.3):

$$\Upsilon = zf(\omega) = \omega f(\omega)\theta = \varphi\theta,$$

$$\varphi = \begin{cases} 10^{\eta-\eta_{\min}} & \text{if } \eta \leq \eta_{\min} \\ 1 & \text{otherwise} \end{cases}. \quad (3.5)$$

where $\varphi$ is the normalized scalar regressor.

***Remark 3.*** *At this step of the research, it is necessary to briefly note that the unknown parameters can be calculated analytically from the regression (3.5) over finite time. Based on the definition of the regressor φ in (3.5), if there is no measurement noise w, the unknown parameters θ can be found over the finite time $t_k$ using the following procedure:*

$$\theta = \Upsilon(t_k),$$
$$\text{IF } \varphi(t) = 1, \text{ THEN } t_k = t. \tag{3.6}$$

*However, as in reality the regression (3.5) inevitably contains some noise components $w(t_k)$, then $\varphi(t_k) = 1$ means $\Upsilon(t_k) = \theta + w(t_k)$. It might also happen that $w(t_k) > 0$ at all time moments $t_k$. Then the estimate (3.5), which is obtained over the finite time interval, is not accurate enough. At the same time, the application of the gradient method to identify the parameters θ of the regression $\Upsilon(t) = \varphi(t)\theta + w(t)$ allows one to obtain the estimation of θ using several time points instead of just one $t = t_k$. This helps to reduce the influence of the disturbances on the identification accuracy at each certain time point $t_k$.*

***Proposition 1.*** *The normalized regressor $\varphi \in [0; 1]$.*

***Proposition 2.*** *If the regressors $\omega_j$ are finitely exciting (2.8) over the time interval $[t_s; t_s+T]$, the following holds for the normalized regressor φ over the time interval $[t_s; t_s+T]$:*

1) *when $\eta \leq \eta_{min}$, it is true that*

$$10^{-2\eta_{min}} \alpha_j \leq \int_{t_s}^{t_s+T} \varphi^2(\tau)d\tau \leq T. \tag{3.7}$$

2) *when $\eta > \eta_{min}$, it is true that*

$$0 < \Delta \leq \int_{t_s}^{t_s+T} \varphi^2(\tau)d\tau = T, \tag{3.8}$$

*where Δ has the same value for all regressors $\omega_j$, which belong to the class (2.10).*

*The proof of both propositions 1 and 2 are shown in Appendix.*

The estimation loop, which is developed using the normalized regression and the gradient descent method, is written as:

$$\dot{\tilde{\theta}}_i(t) = -\gamma\varphi^2\tilde{\theta}_i(t). \tag{3.9}$$

The differential equation (3.9) is solved over the interval $[t_s; t_s+T]$:

$$\tilde{\theta}_i(t_s + T) = e^{-\gamma\int_{t_s}^{t_s+T}\varphi^2(\tau)d\tau}\tilde{\theta}_i(t_s). \tag{3.10}$$

Using Proposition 2, the two important corollaries are obtained for the solution (3.10).

***Corollary 1.*** *When $\eta \leq \eta_{min}$, then the following inequalities hold for the parameter error $\tilde{\theta}_i(t_s + T)$:*

$$e^{-\gamma T}\tilde{\theta}_i(t_s) \leq \tilde{\theta}_i(t_s + T) \leq e^{-\gamma\alpha_j 10^{-2\eta_{min}}}\tilde{\theta}_i(t_s) \tag{3.11}$$

***Corollary 2.*** *When $\eta > \eta_{min}$, then for all regressors, which belong to the class (2.10), the parameter error $\tilde{\theta}_i(t_s + T)$ is bounded from above as:*

$$\tilde{\theta}_i(t_s + T) = e^{-\gamma T}\tilde{\theta}_i(t_s) \leq e^{-\gamma\Delta}\tilde{\theta}_i(t_s). \tag{3.12}$$

In Corollaries 1 and 2, the situations are considered when either the condition $\eta > \eta_{min}$ or $\eta \leq \eta_{min}$ is satisfied over the whole excitation interval for all regressors $\omega_j$. But it is also necessary to pay attention to the situation when $\eta > \eta_{min}$ for $\omega_j$ up to some time point $T_j$, and $\eta \leq \eta_{min}$ is true after $T_j$. For this purpose, the following statement is introduced:

***Proposition 3.*** *Let $\omega_j \in$ FE over the time interval $[t_s; t_s+T]$, and there exists the time point $T_j \in (t_s; t_s+T]$, such that $\forall t \in [t_s; t_s+T_j]$ the inequality $\eta > \eta_{min}$ holds for $\omega_j$, while $\forall t \in (t_s+T_j; t_s+T]$ $\eta \leq \eta_{min}$ is true.*
*Then:*
*1) the following holds for the normalized regressor:*

$$0 < \Delta_{min} \leq \int_{t_s}^{t_s+T} \varphi^2(\tau) d\tau \leq T. \qquad (3.13)$$

where $\Delta_{min}$ *has the same value for all regressors $\omega_j$, which belong to the class* (2.10).

*2) the parameter error $\tilde{\theta}_i(t_s + T)$ is bounded from above for the regressors $\omega_j$ according to:*

$$\tilde{\theta}_i(t_s + T) \leq e^{-\gamma \Delta_{min}} \tilde{\theta}_i(t_s). \qquad (3.14)$$

*The proof of Proposition 3 and the definition of $\Delta_{min}$ is shown in Appendix.*

Thus, it follows from the equations (3.11), (3.12) and (3.14) that the choice of the parameter $\eta_{min}$ (so that to satisfy the inequality $\eta > \eta_{min}$ over the regressor excitation interval $[t_s; t_s+T]$) is necessary and sufficient to limit the parameter error $\tilde{\theta}_i(t_s + T)$ from above to the same value for the regressors, which belong to the class (2.10). So, the same value of the adaptation gain $\gamma$ can be used for all such regressors in the estimation loop (3.9).

***Remark 4.*** *The value of $\eta_{min}$ defines the regressor $\omega_j$ of minimal order of magnitude, for which the normalized regressor $\varphi$ with the normalized excitation* (3.8) *or* (3.13) *can be obtained. In turn, such normalized regressor $\varphi$ defines the value of $\Delta_{min}$ according to* (A7). *Therefore, it is assumed that, if $\eta_{min}$ has already been defined, the proposed normalization of the regressor excitation allows one to separate some subclass of regressors from the class* (2.10), *for which the same upper bound of the parameter error exists and the same value of the parameter $\gamma$ can be used.*

As for the regressors $\omega_j$, which are not included in this subclass, i.e., with an order less than or equal to $\eta_{min}$, *their excitation level* (3.7) *will not be normalized over the whole excitation interval. Therefore, in practice, the value $\eta_{min}$ should be chosen on the basis of a priori data about the minimum possible order of the magnitude of the regressor $\omega$, the maximum possible order of the measurement noise, as well as the maximum possible accuracy of the determinant calculation in* (2.5) *using the numerical procedures. If the parameter $\eta_{min}$ is not chosen properly, such that $\eta \leq \eta_{min}$ $\forall t \in [t_s; t_s+T]$, then, if $\eta_{min} < 0$, the proposed normalization allows one* (3.7) *to increase the initial level of the regressor excitation by a factor of $10^{-2\eta_{min}}$. This proves that it is better to choose the $\eta_{min}$ value from the subset of negative numbers.*

Let the properties of the normalized estimation loop (3.9), which are described in this section of the paper, be demonstrated by an example.

***Example 2.*** *Let $t_s = 0$ s, $T = 10$ s, $\eta_{min} = -2$ and the following regressors be considered: $\omega_1 = e^{-1t}$ and $\omega_2 = 10e^{-1t}$. Firstly, let the time points, when each regressor changes its order of magnitude, be found over the interval $[0; 10]$:*

$$t_1(\eta = -1) = t_2(\eta = 0) = \frac{ln(0,1)}{-1} = \frac{ln\left(\frac{1}{10}\right)}{-1} \approx 2,3;$$

$$t_1(\eta = -2) = t_2(\eta = -1) = \frac{ln(0,01)}{-1} = \frac{ln\left(\frac{0,1}{10}\right)}{-1} \approx 4,61; \qquad (3.15)$$

$$t_2(\eta = -2) \approx 6,91;$$

*Thus, given $\eta_{min} = -2$, the regressors $\varphi_1$ and $\varphi_2$ are written as follows over the corresponding intervals:*

| $t$ | $\varphi_1$ | $t$ | $\varphi_2$ |
|---|---|---|---|
| $[0;4{,}61]$ | 1 | $[0;6{,}91]$ | 1 |
| $[4{,}61;10]$ | $10^{\eta-2} = \|\omega\|10^{-2} = e^{-1(t-4{,}61)}$ | $[6{,}91;10]$ | $10^{\eta-2} = \|\omega\|10^{-2} = e^{-1(t-6{,}91)}$ |

*Then, in terms of Proposition 3, the time moment $T_j$ is 4.61 s for $\varphi_1$ and 6.91 s for $\varphi_2$. Hence, according to (A7), the common $\Delta_{min}$ for both $\varphi_1$ and $\varphi_2$ can be chosen from the interval $(0; 4.61]$.*

*Using the notion of the regressors $\varphi_1$ and $\varphi_2$, the integral from (3.10) is calculated:*

$$\int_0^{10} \varphi_1^2(\tau)d\tau \approx \int_0^{4{,}61} 1^2 d\tau + \int_{4{,}61}^{10} e^{-2(t-4{,}61)} d\tau \approx 5{,}11 \geq \Delta_{min},$$
$$\int_0^{10} \varphi_2^2(\tau)d\tau \approx \int_0^{6{,}91} 1^2 d\tau + \int_{6{,}91}^{10} e^{-2(t-6{,}91)} d\tau \approx 7{,}409 \geq \Delta_{min}. \quad (3.16)$$

*The parameter error for the regressors $\varphi_1$ and $\varphi_2$ over the interval $[0; 10]$ is written as:*

$$\tilde{\theta}_i^{\varphi_1}(10) \approx e^{-\gamma 5{,}11} \tilde{\theta}_i(t_s) \leq e^{-\gamma \Delta_{min}} \tilde{\theta}_i(0),$$
$$\tilde{\theta}_i^{\varphi_2}(10) \approx e^{-\gamma 7{,}409} \tilde{\theta}_i(t_s) \leq e^{-\gamma \Delta_{min}} \tilde{\theta}_i(0). \quad (3.17)$$

*As follows from (3.16) and (3.17), there exists the same upper bound of the parameter errors obtained by the loop (3.9) for different regressors. And the value of this bound is defined by the adaptation rate $\gamma$, which can be chosen as the same value for all regressors.*

## 4. Comparison with the conventionally normalized gradient-based estimation law

In this section of the paper the developed estimation loop will be compared to the already known gradient loop with conventional regressor normalization [15].

The equation of the gradient-based estimation loop with the classical normalization is written as:

$$\dot{\tilde{\theta}}_i = -\gamma \frac{\omega^2}{1+\omega^2} \tilde{\theta}_i. \quad (4.1)$$

If the condition (2.8) is met, then the following equality holds for the normalized regressor:

$$0 < \alpha \leq \int_{t_s}^{t_s+T} \frac{\omega^2}{1+\omega^2}(\tau)d\tau < T. \quad (4.2)$$

In contrast to the classical gradient loop (2.6), the excitation level of the normalized regressor in (4.1) is bounded from above by the value $T$ for any regressor. However, comparing to the normalization approach proposed in this paper, the following disadvantages should be noted. Firstly, the upper bound of the excitation in (4.2) is strictly below $T$, and secondly, the lower bound of the excitation in (4.2) can be an arbitrarily small number $\alpha$. The lower the amplitude $A$ of the regressor $\omega$, the lower the value of $\alpha$.

Given (4.2), the upper bound of the solution of the differential equation (4.1) over the interval $[t_s; t_s+T_1]$ is as follows:

$$\tilde{\theta}_i(t_s + T) \leq e^{-\gamma \alpha} \tilde{\theta}_i(t_s). \quad (4.3)$$

Thus, it follows from (4.2) by analogy with (2.9) that the estimation loop with classical normalization (4.1) also requires adjustment of the adaptation rate $\gamma$ in order to keep the value of γα constant for the regressors with different levels of excitation α. This is not required when the developed estimation loop (3.9) is applied.

## 5. Numerical Example

In Matlab/Simulink let the developed estimation loop (3.9) with normalization of the regressor excitation, the gradient loop (2.6) and the gradient loop (4.1) with the conventionally normalized regressor be compared. The simulation will be conducted using the Euler numerical integration method with a constant step size of $\tau_s = 10^{-2}$ second. The following transfer function was chosen as a plant, which parameters were to be identified:

$$y(t) = \frac{b_1 p + b_0}{p^2 + a_1 p + a_0} u(t) = \frac{2p+1}{p^2 + 1p + 2} u(t). \tag{5.1}$$

The parameters of the filters (2.2), time delay values (2.3) and $\eta_{min}$ value were defined as follows:

$$\psi_1 = 20;\ \psi_0 = 100;\ d_1 = 0,2;\ d_2 = 0,4;\ d_3 = 0,6;\ \eta_{min} = -12. \tag{5.2}$$

The time delay values were chosen according to the recommendations given in [18], and the value of the parameter $\eta_{min}$ was defined using Remark 4.

Taking into consideration Remark 1, the control action signal for the plant (2.1) was chosen to be constant for each certain experiment:

$$u = 1;\ u = 10;\ u = 100. \tag{5.3}$$

The initial value of the parameter error was defined as $\tilde{\theta}(0) = -\theta$. The adaptation gain values for each of the three estimation loops under consideration were chosen as:

$$\gamma = 10^4;\ \gamma_{NE} = 0,1;\ \gamma_{NR} = 10^4. \tag{5.4}$$

where $\gamma$, $\gamma_{NE}$, $\gamma_{NR}$ are adaptation gains of the loops (2.6), (3.9) and (4.1) respectively.

Figure 1 shows the comparison of the normalized regressors φ obtained from the regressor ω for the control action signals of the different amplitude (5.3).

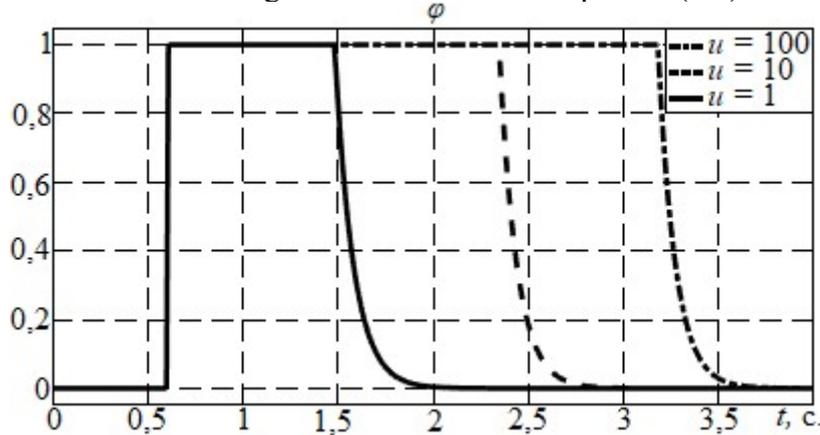

Fig. 1. Comparison of normalized regressors φ for different values of $u$.

The transients shown in Fig.1 confirmed the conclusions made in the propositions and, in terms of Proposition 3, allowed us to find the range of the possible values of Δ as (0; 0.88].

Figure 2 shows a comparison of the transients of $\|\tilde{\theta}\|$, which were obtained using the estimation loops (2.6), (3.9) and (4.1). The plot of the parameter error norm of the developed estimation loop (3.9) also includes its upper bound (UB), which was calculated with (3.14). Δ = 0.7 and $t_s$ = 0 seconds.

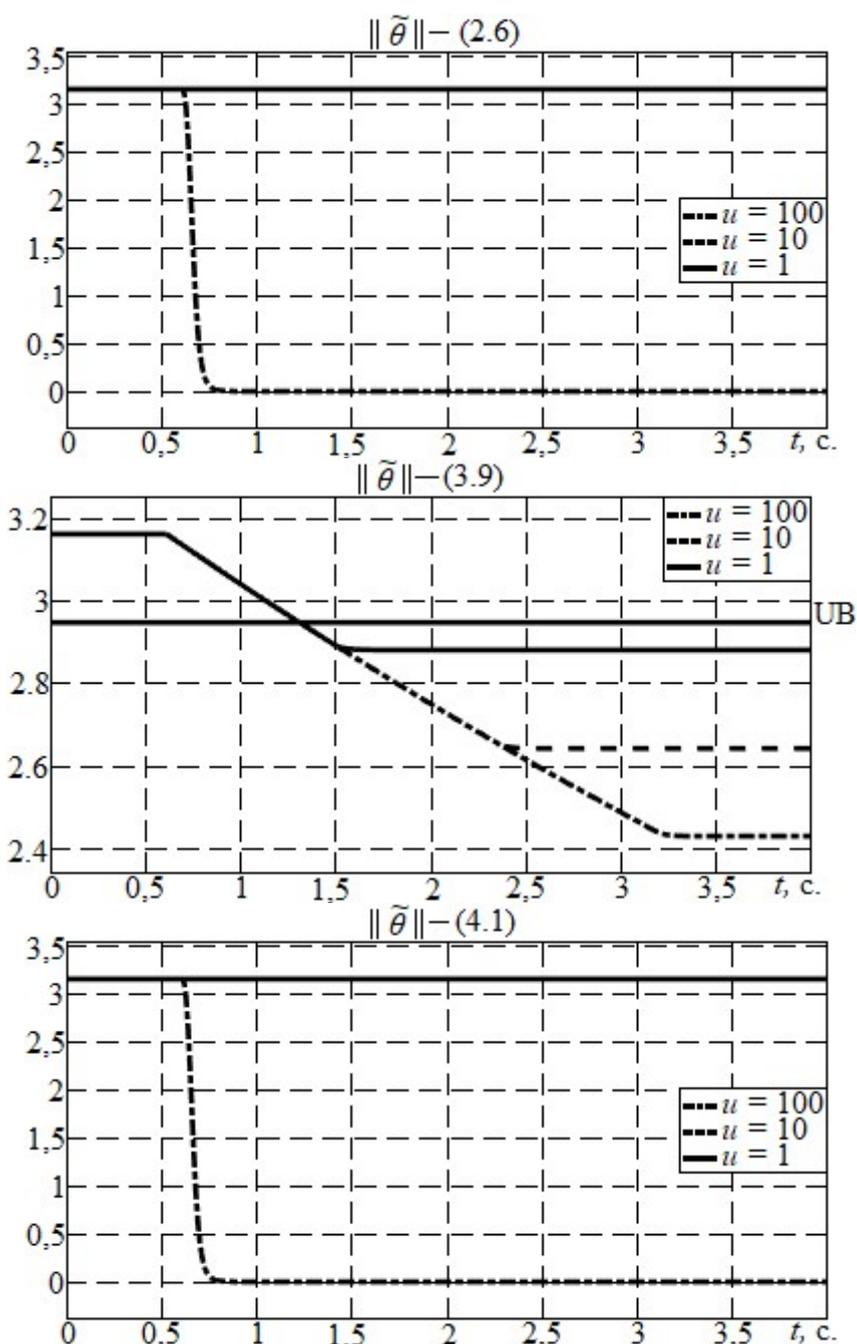

Fig. 2. Norm of parameter errors of estimation loops (2.6), (3.9) and (4.1).

As it follows from Fig.2, in contrast to (2.6) and (4.1), the developed estimation loop prevented situations, when $\|\tilde{\theta}(t \to \infty)\| \to \|\tilde{\theta}(t_0)\|$ for some regressors, whereas $\|\tilde{\theta}(t \to \infty)\| \to 0$ for other ones. This is the main result of this research.

Next, the developed estimation loop was modeled using different values of the control action (5.3) and the adaptation rate.

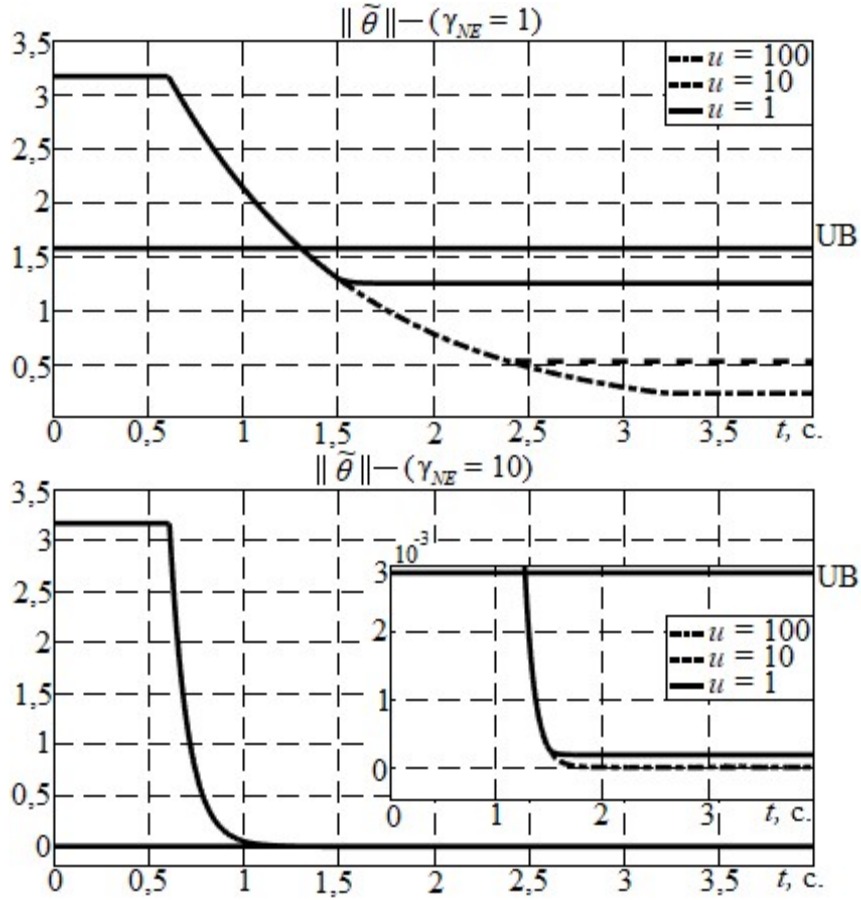

Fig. 3. Parameter error norm of estimation loop (3.9) for different values of adaptation rate γ.

The results of this experiment showed that the value of the parameter γ in (3.9) allowed one to adjust the upper bound of the parameter error and avoid situations when $\left\|\tilde{\theta}(t \to \infty)\right\| \to \left\|\tilde{\theta}(t_0)\right\|$ for some regressors, and $\left\|\tilde{\theta}(t \to \infty)\right\| \to 0$ for others.

### 6. Conclusion

In order to identify the parameters of the linear dynamic plants, which functioned in various operation modes, under the condition that the adaptation rate of the estimation loop had been chosen once and then kept constant, the procedure of normalization of the regressor excitation was developed in this study. It could be useful for the identification of models of industrial plants functioning under the condition of both the step-like parameters and setpoint values change.

Since the effect of the measurement noise on the obtained results was noted briefly only (see Remarks 3 and 4), it is planned to investigate the properties of the developed normalization procedure in more detail under such conditions in future research. Also, in the following studies it is planned to use the proposed approach for normalization of the excitation of the integral-based regressor in the procedure [19].

**APPENDIX**

***Proof of Proposition 1.*** When $\eta > \eta_{min}$, the regressor φ = 1 according to (3.5). Otherwise $\eta - \eta_{min} \leq 0$, whence it follows that $10^{\eta-\eta_{min}} \leq 10^0$. Hence φ belongs to the interval [0; 1], as was to be proved.

***Proof of Proposition 2.*** Taking into consideration (3.1)-(3.5), the equation of the regressor φ is rewritten as:

$$\varphi = 10^{\eta - sat(\eta)} = |\omega| 10^{-sat(\eta)}. \tag{A1}$$

The expression for $\omega^2$ is found from (A1) and substituted to (2.10):

$$\int_{t_s}^{t_s+T} \omega_j^2(\tau) d\tau = 10^{sat(\eta)} \int_{t_s}^{t_s+T} \varphi^2(\tau) d\tau \geq \alpha_j. \tag{A2}$$

So, considering (3.4), when $\eta \leq \eta_{min}$, the proof of the lower bound from (3.7) is immediate. Also, applying the fundamental theorem of calculus and considering $\varphi \in [0; 1]$, the upper bound from (3.7) is obtained from (A2).

When $\eta > \eta_{min}$, the equation (A2) is written as:

$$10^{-\eta} \int_{t_s}^{t_s+T} \omega_j^2(\tau) d\tau = \int_{t_s}^{t_s+T} \varphi^2(\tau) d\tau \tag{A3}$$

Using the Leibniz formula and considering $\varphi = 1$ when $\eta > \eta_{min}$, the following inequality is obtained from (A3):

$$10^{-\eta} \int_{t_s}^{t_s+T} \omega_j^2(\tau) d\tau = \int_{t_s}^{t_s+T} \varphi^2(\tau) d\tau = T \tag{A4}$$

As $T > 0$ according to the finite excitation condition, then there always exists some $0 < \Delta \leq T$ in (A4), such that the inequality (3.8) holds. Since $T$ is the same for all regressors from the class (2.10), the value $\Delta$ can also be assumed to be the same for all regressors (2.10), which completes the proof of the proposition.

***Proof of Proposition 3.*** The finite excitation condition of the regressor φ can be rewritten as follows:

$$\int_{t_s}^{t_s+T} \varphi^2(\tau) d\tau = \int_{t_s}^{t_s+T_j} \varphi^2(\tau) d\tau + \int_{t_s+T_j}^{t_s+T} \varphi^2(\tau) d\tau \geq \int_{t_s}^{t_s+T_j} \varphi^2(\tau) d\tau. \tag{A5}$$

As $\eta > \eta_{min}$ over the interval $[t_s; t_s+T_j]$ and using the proved Proposition 2, the lower bound of (A5) is:

$$\int_{t_s}^{t_s+T_j} \varphi^2(\tau) d\tau = T_j. \tag{A6}$$

Since, according to the proposition being proved, $T_j$ can take values from the interval $(t_s; t_s+T]$, then there exists a moment of time $\Delta_{min}$ such that:

$$0 < \Delta_{min} \leq \min_{j \geq 0} \{T_j - t_s\} \leq T_j - t_s. \tag{A7}$$

Given inequalities (A5), (A6) and (A7), the lower bound is written in the form of (3.13). To obtain the upper bound from (3.13), let, taking into account Proposition 2, the upper bound of the second term from (A5) be written:

$$\int_{t_s+T_j}^{t_s+T} \varphi^2(\tau) d\tau \leq T - T_j. \tag{A8}$$

Adding the upper bounds (A6) and (A8), the upper bound from (3.13) is obtained, which together with the lower bound from (A5) allows us to write the inequality (3.13) in its full form. Using (3.13), it is not difficult to obtain (3.14) for the solution (3.10), which completes the proof of the proposition.